\documentclass[prl,aps,twocolumn,amsfonts,showpacs,groupedaddress]{revtex4-1}
\usepackage{graphicx}
\usepackage{xcolor}
\usepackage{epsfig}
\usepackage{epstopdf}
\usepackage{rotating}
\usepackage[mathscr]{eucal}
\usepackage{amsmath,amssymb,bm,graphics}
\usepackage[pdftex,letterpaper=true]{hyperref}

\let\mathbf=\bm

\begin{document}
\def\Nfour{\mathcal N\,{=}\,4}
\def\Ntwo{\mathcal N\,{=}\,2}
\def\Nc{N_{\rm c}}
\def\Nf{N_{\rm f}}
\def\x{\mathbf x}
\def\q{\mathbf q}
\def\f{\mathbf f}
\def\v{\mathbf v}
\def\C{\mathcal C}
\def\w{\omega}
\def\vs{v_{\rm s}}
\def\S{\mathcal S}
\def\half{{\textstyle \frac 12}}
\def\twothirds{{\textstyle \frac 23}}
\def\third{{\textstyle \frac 13}}
\def\t{\mathbf{t}}
\def\T{\mathcal {T}}
\def\O{\mathcal{O}}
\def\E{\mathcal{E}}
\def\p{\mathcal{P}}
\def\H{\mathcal{H}}
\def\uh{u_h}
\def\R{\ell}
\def\Ro{\chi}
\def\del{\nabla}
\def\eps{\hat \epsilon}
\def\nn{\nonumber}
\def\K{\mathcal K}
\def\inf{\epsilon}
\def\cs{c_{\rm s}}
\def\A{\mathcal{A}}
\def\e{{e}}
\def\r{{\xi}}
\def\x{{\mathbf x}}
\def\w{{w}}
\def\rr{{\xi}}
\def\uo{{u_*}}
\def\u{{\mathcal U}}
\def\G{\mathcal{G}}
\def\Deltax{\Delta x_{\rm max}}
\def\R{\lambda}

\title
{Vortex annihilation and inverse cascades in two dimensional superfluid turbulence}

\author{Paul~M.~Chesler}
\email	{pchesler@physics.harvard.edu}
\author{Andrew~Lucas}
\email	{lucas@fas.harvard.edu}

\affiliation{Department of Physics, Harvard University, Cambridge MA 02138, USA}

\date{\today}

\begin{abstract}
We study two dimensional superfluid turbulence by employing an effective description 
valid in the limit where the density of superfluid vortices is parametrically small.  At sufficiently low temperatures 
the effective description yields an inverse cascade with Kolmogorov energy spectrum $E(k) \sim k^{-5/3}$.  Denoting the number of vortices 
as a function of time by $N(t)$, we find that the vortex annihilation rate scales like $\dot N \sim  N^{5/3}$ in states with an inverse cascade and $\dot N \sim N^2$ for laminar flow.
\end{abstract}


\maketitle

{\it Introduction.---}Two-dimensional turbulent superfluids are now studied experimentally 
in cold atomic gases \cite{neely, kwon}.  
An important open question is the direction of energy cascades in superfluid turbulence.    
At temperatures of order the critical temperature, where dissipative effects are large, holographic duality
has yielded evidence for direct cascades  \cite{chesler}, where energy injected at large length scales is dissipated at small scales.     In contrast, studies using weakly damped or zero temperature 
Gross-Pitaevskii equations suggest inverse cascades \cite{reeves, billam, simula}, analogous to classical fluids, where short distance forcing creates large vortex structures.
 Understanding these differences and their physical origin is of great interest.  

Superfluid vortices are discrete with quantized circulation and opposite circulation
vortices can annihilate.   A simple observable is the vortex annihilation rate $\dot N \equiv dN/dt$ 
with $N(t)$ the number of vortices as a function of time $t$.  
Does the nature of the superfluid flow imprint itself on the vortex annihilation rate?
Is the annihilate rate different in turbulent states, and/or sensitive to the direction of the cascade?

Annihilation events can be driven by vortex-vortex interactions mediated by sound
or by thermal effects such as vortex drag \cite{boris}.   
In this letter we focus on vortex annihilation in the limit of an asymptotically dilute system of vortices.
In this limit vortex-vortex interactions mediated by sound are negligible \cite{Lucas:2014tka}
and vortex annihilation is driven by thermal effects.  

We employ an effective description valid in the limit where the average vortex separation 
$\lambda$ is much larger than the vortex core size $\xi$.
In the effective description dissipation is encoded in a 
phenomenological temperature-dependent parameter $\hat \eta(T)$,
which is the ratio of the strengths of drag and Magnus forces on a vortex.

In the limit of small dissipation ($\hat \eta \ll 1$), 
via numerical simulations
we demonstrate the system 
can enjoy an inverse cascade, with macroscopic clusters of like-circulation vortices forming
and the energy spectrum obeying Kolmogorov's $k^{-5/3}$ scaling.  
Indeed, for $\hat \eta = 0$ the effective dynamics reduces to 
two dimensional classical hydrodynamics, where the existence of inverse cascades with Kolmogorov scaling is well known.
Our main result is that when $\hat\eta \ll 1$, the inverse cascade imprints itself 
on the vortex annihilation rate: in states with inverse cascades $\dot N \sim - N^{5/3}$
whereas for laminar flow $\dot N \sim -  N^{2}$.  
These different $N$ scalings can be used to distinguish states with inverse cascades from those without 
and may be easier to access experimentally than other probes of turbulence such as the energy spectrum.

{\it Effective dynamics.---}We wish to study the dynamics of a finite temperature state with $N$ superfluid vortices.
Denoting the vortex trajectories by $\bm X_q(t)$ with $q = 1,2,\dots, N$, we aim to find
an equation of motion for the $\bm X_q(t)$ in the limit where the average vortex separation $\R \to \infty$.
The effective description we construct below is simply the first order HVI equations \cite{Hall18121956,1964AnPhy..29..335I, halperin1}.

Within the framework of effective field theory, the dynamics of $\bm X_q$ can be obtained by ``integrating out"
all degrees of freedom except the vortex positions (see for example \cite{Endlich:2010hf,Endlich:2013dma,Lucas:2014tka}).  
We choose to focus on the leading order equations of motion for $\bm X_q(t)$ in the $\R\to \infty$ limit.
The leading order equations of motion  follow from momentum conservation and Galilean invariance.

Consider  the motion of a  vortex with trajectory $\bm X(t)$ in  
constant background super and normal fluid velocity fields $\bm V$ and $\bm U$, respectively.
We restrict ourselves to the limit where $\bm V$ and $\bm U$ 
and $\dot {\bm X}$ are asymptotically small 
and specialize to the case where  
the vortex has unit circulation $\kappa = \pm 1$.
Denoting the quantized unit of vorticity as $\Omega$ and the local superfluid velocity by $\bm v$, this means
$\oint d \bm \ell \cdot \bm v = \kappa \Omega$
for every contour which encloses the vortex.

The relative motion of the vortex to the normal flow will in general result in a transfer of momentum
$d \bm p_{\rm s \to \rm n}/dt$
from the super flow to the normal flow.   
In the limit $\dot {\bm X} , \ \bm U \to 0$ the most general form of $d \bm p_{\rm s \to \rm n}/dt$ consistent with Galilean invariance
is %
\footnote
  {
  Additional 
  forces, such as stochastic forces due to thermal fluctations, can also be
  added to (\ref{eq:generalmomentumloss}).  In holographic models of 
  superfluid turbulence like that explored in \cite{chesler}, stochastic 
  forces are suppressed.  We choose to ignore such terms.
  }
\begin{equation}
\label{eq:generalmomentumloss}
\frac{d p^i_{\rm s \to \rm n}}{dt} =  -\eta (\dot X^i - U^i) - \eta' \kappa \Omega  \epsilon^{ij}  (\dot X^j - U^j),
\end{equation}
where $\epsilon^{ij}$ is the antisymmetric symbol with $\epsilon^{12} = 1$.
Here and in what follows we adopt the convention of implicit summation
over repeated spatial indices (superscripts).
The phenomenological constants $\eta$ and $\eta'$  depend 
on microscopic physics and must vanish at zero temperature
where the normal component of the system vanishes.  

Momentum conservation demands the momentum transfer rate (\ref{eq:generalmomentumloss})
be balanced by flux of momentum $d \bm p_{\infty \to \rm core}/dt$ carried 
by the super flow from large distances to the vortex core. 
In the limit $\bm V, \bm U,  \dot {\bm X} \to 0$, where the system becomes static,   
the flux of momentum through any surface $\Sigma$ must be independent of $\Sigma$.
Hence, one can evaluate $d \bm p_{\infty \to \rm core}/dt$ in the asymptotic far zone where the dynamics 
can be treated with hydrodynamics.
Straightforward calculations using superfluid hydrodynamics yield (see for example \cite{PhysRevB.55.485})
\begin{align}
\label{eq:momentumcons}
\frac{d  p^i_{\infty \to \rm core}}{dt} = \rho_s \kappa \Omega \epsilon^{ij} (\dot X^j - V^j), 
\end{align}
where $\rho_s$ is the ambient superfluid density.  Eq.~(\ref{eq:momentumcons})
is the standard expression for the Magnus force on a vortex moving through a background superfluid velocity field $\bm V$.  

Equating the momentum loss rate (\ref{eq:generalmomentumloss}) with the flux (\ref{eq:momentumcons}) we obtain the 
single-vortex equation of motion
\begin{equation}
\label{eq:vortexeq}
\rho_s \kappa \Omega \epsilon^{ij} ( \dot X^j - V^j) = 
-\eta (\dot X^i - U^i) - \eta' \kappa \Omega  \epsilon^{ij}  (\dot X^j - U^j).
\end{equation}
This is simply the first order HVI equation. 

With the single vortex equation of motion (\ref{eq:vortexeq}) it is a simple matter 
to construct the equations of motion for an asymptotically dilute system of vortices interacting with each other.
Each vortex will source super and normal fluid velocity fields
which decay like $1/d$ with $d$ being the distance to the vortex core.
In the $\R \to \infty$ limit non-linear interactions between flows produced by neighboring vortices are $1/\R$ 
suppressed and the net super and normal velocity fields are simply the linear superposition 
of those generated by individual vortices.
Therefore, in the $\R\to \infty$ limit the dynamics of a vortex at $\bm X_q$ must be governed by 
(\ref{eq:vortexeq}) with the ``background" fields $\bm V$ and $\bm U$  being the sum of the 
super and normal fluid velocity fields created by all the other vortices far away from $\bm X_q$.   
By virtue of the $1/d$ decay of the flows produced by each vortex, 
the resulting ``background" fields will be both parametrically small and slowly varying in the neighborhood of each vortex.

In equilibrium the superfluid velocity field of a single vortex of winding number $\kappa$ is $\kappa \bm v_{\rm vortex}(\bm x)$ with
\begin{equation}
\label{eq:vortexvel}
v_{\rm vortex}^i(\bm x) \equiv - \frac{\Omega}{2\pi} \frac{ \epsilon^{ij} x^j}{x^2}.
\end{equation}
Hence in the $\R\to \infty$ limit the background superfluid velocity a vortex at $\bm X_m$ moves in is 
\begin{equation}
\label{eq:localvelfelt}
\bm V_m(t) \equiv \sum_{m \neq q} \kappa_q \bm v_{\rm vortex}(\bm X_m(t) - \bm X_q(t)).
\end{equation}

Likewise, via (\ref{eq:generalmomentumloss}) each vortex deposits momentum in the normal flow 
at rate $\sim \dot {\bm X_q}$.  Via (\ref{eq:localvelfelt}) $\bm V_m \sim 1/\R$ so $\dot {\bm X}_q \sim 1/\R$.
Since the normal velocity field decays like an inverse power of distance, it follows that vortex interactions 
mediated by normal flow are $1/\R$ suppressed relative to the that mediated by the super flow (\ref{eq:localvelfelt})
and the normal flow can be taken to be its background value.  

Therefore, to obtain the equations of motion for vortex at $\bm X_m$, in Eq.~(\ref{eq:vortexeq}) we  set 
$\bm U = 0$ and replace $\bm V \to \bm V_m$, $\bm X \to \bm X_m$.
Under the rescaling $t \to t/[ (1 + \eta'/\rho_s)(1 + \hat \eta^2)]$ where
the dimensionless drag coefficient $\hat \eta$ is
\begin{align}
 \textstyle \hat \eta \equiv \frac{\eta}{\Omega (\rho_s  + \eta')},
\end{align} 
the resulting equations of motion are simply \cite{Hall18121956,1964AnPhy..29..335I, halperin1}
\begin{equation}
\label{eq:vortexeq3}
\dot X_m^i = \textstyle ( \delta^{ij} + \hat \eta \kappa_m \epsilon^{ij}) V_m^j.
\end{equation}
For any $s$ these equations  are invariant under  
\begin{align}
\label{eq:rescalings}
& t \to s^2 t, & \bm X_q \to s \bm X_q.
\end{align}

Numerical simulation of  (\ref{eq:vortexeq3}) with $\hat \eta = 0$
can yield inverse cascades with Kolmogorov energy spectrum \cite{siggia}.
With $\hat \eta = 0$ the equations of motion reduce
to point vortex dynamics where vortices move at the local superfluid velocity
and vortex annihilation is forbidden \cite{novikov}.
However, when $\hat \eta > 0$  vortex velocities have a component $\hat \eta \epsilon^{ij} V_m^j$ orthogonal to the local superfluid velocity.  
This results in repulsive interactions between like-circulation vortices and attractive interactions between opposite circulation vortices.
When $\hat \eta > 0$  vortex annihilation is allowed: pairs of opposite circulation vortices can come together to the same point in a finite time!

To study vortex annihilation in states with inverse cascades we solve the equations of motion numerically at finite $\hat \eta$.
However, the equations of motion (\ref{eq:vortexeq3}) become singular when vortices annihilate.
Indeed, the effective description 
breaks down when the vortex separation  is comparable to the vortex core size $\xi$.
However, as pairs of opposite 
winding number come together their combined velocity fields nearly cancel
and their influence on all the other vortices becomes negligible \textit{before} they 
annihilate. 
To avoid singularities in our numerical simulations arising when vortices annihilate, 
when two vortices of opposite winding 
number come within a set distance $\xi$, we simply delete the 
pair.  We choose units where $\xi = 1$.

We solve the equations of motion (\ref{eq:vortexeq3}) numerically
in a periodic box of size $L$ \cite{stremler}.  We set $\Omega = 2 \pi$ and 
choose initial conditions with
$\kappa_m = {\rm sgn}  ( \sin Q x_m  )$,  
and $\sum \kappa_m=0$.
We employ two sets of parameters.  For the first set of parameters we choose
$L = 100$, $Q = 8 \pi/L$ and $\hat \eta = 2.5 \times 10^{-3}$
and $N = 640$.  We refer to these 
parameters and initial conditions as the ``small box"  parameters.  
For the second set, $L = 1000$, $Q = 20 \pi/L$ and $\hat \eta = 2 \times 10^{-2}$
and $N = 4000$.  We refer to these parameters and initial
conditions as the ``large box" parameters.

\begin{figure*}[ht]
\includegraphics[scale = 0.55]{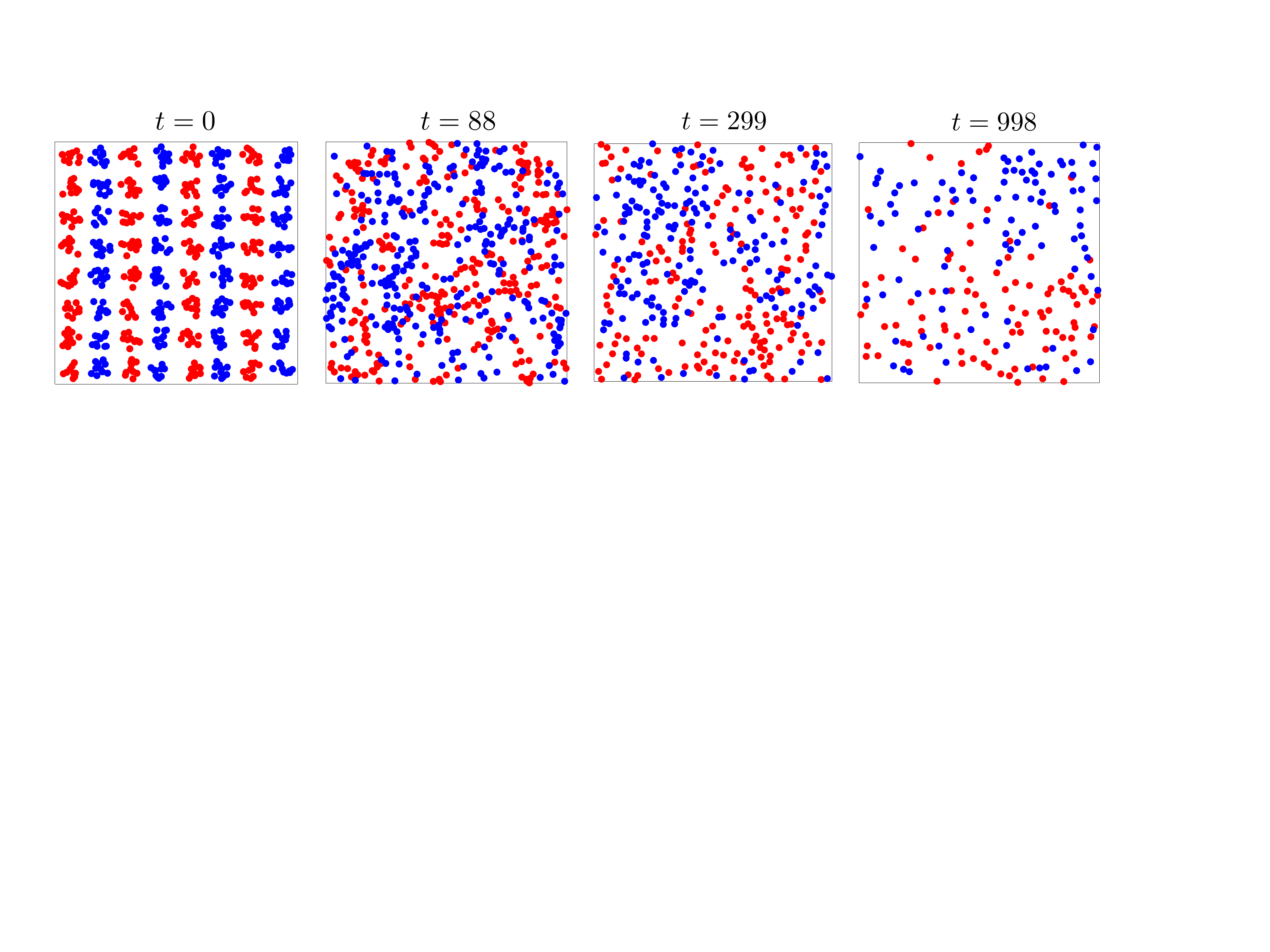}
\caption{The vortex positions at four times for small-box parameters.   The red dots in the plots denote the position of $\kappa = +1$ vortices
and the blue dots denote the position of $\kappa = - 1$ vortices.  Note the formation of 
clusters of vortices of like-winding number vortices.  By time $t = 998$ the system 
consists of one cluster of $\kappa = -1$ vortices (upper half plane) and one cluster 
of $\kappa = +1$ vortices (lower half plane).  Due to vortex annihilation events the number of vortices decreases as a function of time.
\label{fig:snapshots}
} 
\end{figure*}

{\it Inverse cascades.---}
Fig.~\ref{fig:snapshots} shows the evolution of a system of vortices with small-box parameters 
at times $t = 0$, $t = 88$, $t = 299$ and $t = 998$.  The red dots show the position of $\kappa = +1$ vortices
and the blue dots show the position of $\kappa = - 1$ vortices.  During times $t = 0$ through $t = O(20)$ our system experiences an 
instability which drives the initial state into turbulent evolution.  At all times $t > 0$ shown in Fig.~\ref{fig:snapshots} there is no sign of the initial sinusoidal 
structure present in the initial data at $t = 0$.  

At time $t = 88$ the system consists of many clusters of same winding number vortices with clusters 
ranging from a few vortices to a few dozen vortices.
The clusters rotate collectively with $\kappa = + 1$ clusters
rotating counter clockwise and $\kappa = -1$ clusters rotating clockwise.   By  $t = 299$, the clusters are noticeably larger 
in spatial extent while the total number of clusters has decreased.  
In the upper left quadrant there is a large cluster of $\kappa = -1$ vortices and in the lower right quadrant 
there is a large cluster of $\kappa = +1$ vortices.  By  $t = 998$ the system consists of two well-defined large clusters of vortices.  The upper
half plane mostly contains $\kappa = -1$ vortices and the lower half plane mostly contains $\kappa = +1$ vortices.  
This \textit{self sorting} feature --- where vortices of like vorticity tend to clump together to produce larger and larger 
vortex clusters --- is a tell tale signature of an inverse cascade.   

\begin{figure}[h]
\includegraphics[scale = 0.33]{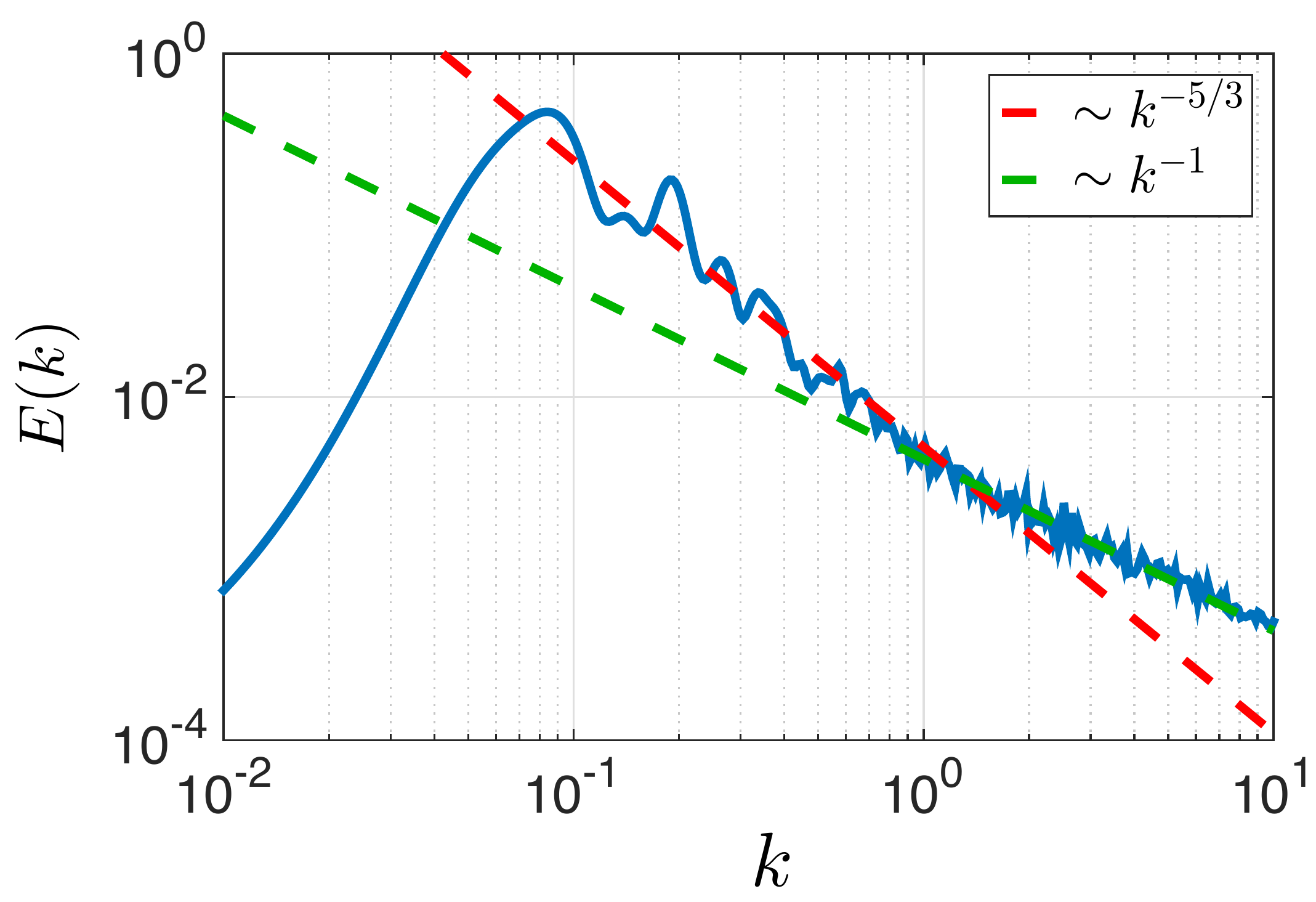}
\caption{The power spectrum of the superfluid velocity at time $t = 88$ for small-box parameters.
The Kolmogorov scaling (\ref{eq:kolscaling}) applies over $k \in (10^{-1},10^{0})$.  
\label{fig:kolscaling}
} 
\end{figure}

To further support the identification of an inverse cascade, we consider the superfluid velocity power spectrum
\begin{equation}
\label{eq:spectrumdef}
E(k) \equiv \frac{1}{L^2} \frac{\partial}{\partial k} \int_{k' \leq k} \frac{d^2 k'}{(2 \pi)^2} | \bm v(\bm k') |^2,
\end{equation}
with $\bm v(\bm k)$ the Fourier transform of the superfluid velocity field.  According to Kolmogorov's 
1941 theory of classical turbulence, physics at scale $k$ --- within an inertial range $(\Lambda_-,\Lambda_+)$ ---
only depends on $k$ and the 
rate $\varepsilon$ in which the energy per unit mass is cascaded from mode to mode.
Dimensional analysis then fixes 
\begin{equation}
\label{eq:kolscaling}
E(k) \sim \varepsilon^{2/3} k^{-5/3},
\end{equation}
In the limit of an asymptotically dilute system of vortices the superfluid velocity is 
\begin{equation}
\label{eq:supervelocityfield}
\bm v(t,\bm x) = \sum_m \kappa_m \bm v_{\rm vortex}(\bm x - \bm X_m(t)).
\end{equation}
In Fig.~\ref{fig:kolscaling} we plot
$E(k)$ at time $t = 88$.    Our numerical results are consistent with the Kolmogorov scaling (\ref{eq:kolscaling})
in the inertial range $\Lambda_- \approx 10^{-1}$ and $\Lambda_+ \approx 10^0$.  For $k > \Lambda_+$ 
our numerical results are consistent with the scaling $E(k) \sim k^{-1}$.  On dimensional grounds 
the power spectrum of the single vortex velocity field (\ref{eq:vortexvel}) scales like
$E(k) \sim k^{-1}$.  Therefore, the knee at $k = \Lambda_+$ 
denotes the transition from the collective physics of interacting vortex clusters to single-vortex physics.  
We note $\Lambda_- \sim 2\pi /L$.

In the limit $\hat \eta \to 0$ it is natural that the dynamics yield an inverse cascade.
For $\hat \eta = 0$ the equations of 
motion (\ref{eq:vortexeq3}) imply that the superfluid velocity field (\ref{eq:supervelocityfield}) satisfies the 
two dimensional Euler equation
\begin{equation}
\label{eq:euler}
\partial_t \omega + \bm v \cdot \del \omega = 0,
\end{equation}
where $\omega\equiv \epsilon^{ij} \del^i v^j = \sum_q \kappa_q \Omega \delta^2(\bm x - \bm X_q(t))$ is the vorticity.
Hence, in the limit $N \to \infty$ with $\hat \eta \to 0$ it is natural to expect 
the macroscopic dynamics to indistinguishable from classical fluid dynamics \cite{siggia, tabeling}.
It is well known that two dimensional classical turbulence enjoys an inverse cascade 
with Kolmogorov scaling.

\begin{figure}[h]
\includegraphics[scale = 0.32]{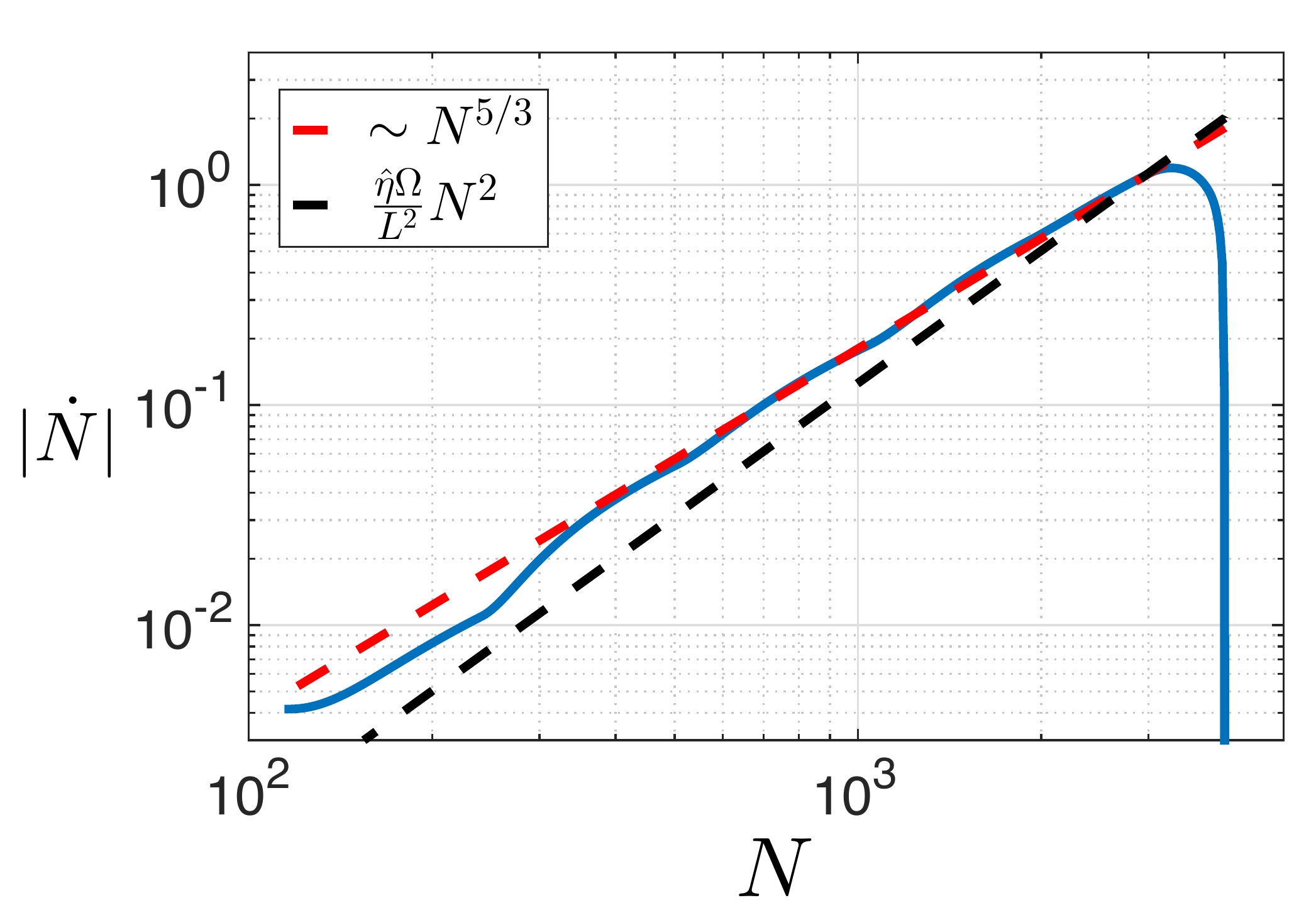}
\caption{The vortex annihilation rate as a function of the number of vortices for large-box parameters.  
$\dot N$ starts off small at early times (large $N$) but increases as vortices mix, as in Fig.~\ref{fig:snapshots}, 
and settles down onto the $\dot N \sim N^{5/3}$ scaling for $N \in (300,3000)$.  Also shown for comparison is Eq.~(\ref{eq:NdotScaling2}), which describes
vortex annihilation for laminar flow.
\label{fig:annrate}
} 
\end{figure}

\textit{Vortex annihilation rate.---}From Fig.~\ref{fig:snapshots} it is obvious
that the number of vortices is decreasing as a function of time.  
To obtain superior vortex annihilation statistics  
we increase the box size, initial number of vortices, and increase vortex drag by employing the 
large-box parameters.  These initial conditions and parameters also lead to an inverse 
cascade and Kolmogorov scaling (\ref{eq:kolscaling}), albeit in a much larger system and with a higher vortex annihilation rate.
We constructed an ensemble of solutions to the equations of motion (\ref{eq:vortexeq3}) and computed the average vortex annihilation rate
$\dot N$.   

In Fig.~\ref{fig:annrate} we plot $\dot N$ as a function of the number of vortices $N$.
Since the number of vortices decreases as a function of time, early times correspond to large
$N$ and late times correspond to small $N$.
For $N = 4000$, corresponding to $t = 0$, when 
the $\kappa = \pm 1$ vortices are not well-mixed, 
the annihilation rate vanishes.
However, as time progresses vortices mix,
the annihilation rate increases, and  settles onto approximate power law behavior.  Our numerical results 
are consistent with $\dot N \sim N^{5/3}$
for $N \in (N_-,N_+)$
with $N_- \approx 300$ and $N_+ \approx 3000$.

To understand the origin of the $\dot N \sim N^{5/3}$ scaling we focus on the limit $\hat \eta \ll 1$, where the dynamics are approximately 
governed by the Euler equation (\ref{eq:euler}), and treat vortex annihilation perturbatively.  We 
furthermore specialize to states consisting of macroscopically large clusters of vortices.
In this limit the vorticity $\omega$ can be approximated
as continuous and vortex annihilation must occur near $\omega = 0$ contours where opposite circulation 
vortices mix.  The annihilation rate must equal the twice the flux of $\pm 1$ circulation vortices into $\mp 1$ circulation clusters.
The flux of such vortices is 
$\frac{1}{\Omega} \oint_C ds \, \hat r \cdot \bm J_{\rm loc}$ where $\bm J_{\rm loc}$ is the vorticity current in the local rest frame of $\omega = 0$ contours and 
$\hat r = \del \omega/| \del \omega|$ is the normal to $\omega = 0$
contours.  The contour $C$ consists of the segments of $\omega = 0$ contours where $\hat r \cdot \bm J_{\rm loc} < 0$ (\textit{i.e.} where $\pm 1$ circulation vortices flow into $\mp 1$ circulation clusters).

In terms of microscopic variables the vorticity current is $J^i = \sum_q \Omega \kappa_q  \dot {X}^i_q  \delta^2 (\bm x - \bm X_q)$.
Using the equations of motion (\ref{eq:vortexeq3}) and taking the macroscopic limit, this becomes
$J^i = \omega v^i +  n \Omega \hat \eta \epsilon^{ij} v^j$ where $n$ is the density of vortices.
At leading order in $\hat \eta \ll 1$ the local rest frame of $\omega = 0$ contours is governed by the Euler equation
from which we obtain $J^i_{\rm loc} = n \Omega \hat \eta \epsilon^{ij} v^j_{||}$ where 
$\bm v_{||} = \bm v - (\hat r \cdot \bm v) \hat r$ is the velocity tangent to $\omega = 0$ contours.  We thus secure 
\begin{equation}
\label{eq:annrate1}
\dot N = -2 \hat \eta n \oint_{C}  \bm v \cdot d\bm \ell.
\end{equation}

First, consider  vortex annihilation  in laminar flows where
$\omega = 0$ contours are smooth and
$C$ approximately encloses positive circulation clusters.
For systems with zero net vorticity Stokes' theorem then fixes 
$\oint_{C} \bm v \cdot d\bm \ell = \Omega N/2$ .   Using $n = N/L^2$ the annihilation rate (\ref{eq:annrate1})
reads 
\begin{equation}
\label{eq:NdotScaling2}
\dot N = - \frac{\Omega \hat \eta}{L^2} N^2.
\end{equation}
We have tested (\ref{eq:NdotScaling2}) 
in states with laminar flow and found good agreement.
The scaling $\dot N \sim - \hat \eta N^2$ also holds for disordered systems of vortices with no macroscopic clusters, albeit 
with a different coefficient than in (\ref{eq:NdotScaling2}).  We elaborate on this in a forthcoming paper.
For comparison we also plot  (\ref{eq:NdotScaling2}) in Fig.~\ref{fig:annrate}.  
Evidently, the laminar flow annihilation rate is smaller than that in a turbulent state with 
an inverse cascade. 

We note that recent experiments  \cite{kwon} studying vortex annihilation in 
a trapped atomic gas have yielded $\dot N \sim -T^2 N^2$
with $T$ the temperature.  This suggests  $\hat \eta \sim T^2$.  Indeed,
recent calculations using holographic duality yield the low temperature scaling $\hat \eta \sim T^2$ \cite{Dias:2013bwa}.

We now consider vortex annihilation in a  state with an inverse cascade.
In classical two dimensional turbulence $\omega = 0$ contours are fractal with fractal
dimension $4/3$ \cite{bernard2006conformal}.  In the limit $\hat \eta \ll 1$, where the 
superfluid dynamics are approximately governed by the Euler equation (\ref{eq:euler}), it is natural to expect the macroscopic 
structure of $\omega = 0$ contours to also be fractal.
Irregularity of $\omega = 0$ contours 
implies $\hat r \cdot \bm J_{\rm loc}$ oscillates in sign.  Hence  
vortex annihilation doesn't happen everywhere along $\omega = 0$ contours, $C$ is not continuous  and
the integral in (\ref{eq:annrate1}) is not constrained by Stokes' theorem.

Consider annihilation on the boundary of a  vortex cluster of size $R$.
From  Kolmogorov's theory we estimate 
\begin{equation}
\oint_{C} d \bm \ell \cdot \bm v \sim  \varepsilon^{1/3} R^{4/3}.
\end{equation} 
The $R^{4/3}$ scaling encodes that $\omega =0$ contours have fractal
dimension $4/3$ \cite{bernard2006conformal}.  
For quantized vortices $\varepsilon$ can 
only depend on $\Omega$ and $n$.
Under the rescaling symmetry (\ref{eq:rescalings})
the Kolmogorov scaling (\ref{eq:kolscaling}) requires
$\varepsilon \to s^{-4} \varepsilon$.  
Indeed, dimensional analysis fixes $\varepsilon \sim \Omega^3 n^2$.  
Substituting into (\ref{eq:annrate1}) we therefore conclude
\begin{equation}
\label{eq:NdotScaling}
\dot N \sim - \hat \eta N^{5/3},
\end{equation}
which confirms our numerical results seen in Fig.~\ref{fig:annrate}.
Simply put, the $\dot N \sim N^{5/3}$ scaling distinguishes states with inverse cascades from those with laminar flow.  

\textit{Outlook.---}While we have focused on the $\hat \eta \ll 1$ limit in this paper, 
it would be interesting to explore the nature of the dynamics as $\hat \eta$ increases and dissipation 
becomes stronger.  Since finite $\hat \eta$  gives rise to repulsive interactions between like-circulation 
vortices, as $\hat \eta$ increases like-winding number vortices repel, vortex cluster formation is 
degraded and inverse cascades must eventually become impossible.   At larger $\hat \eta$ do turbulent superfluids generically transition to the direct cascade observed in \cite{chesler}?  If so, is there a signature
in the vortex annihilation rate?  We leave these interesting questions for future work.

\textit{Acknowledgements.---}We would like to thank Achim Rosch, Subir Sachdev, Yong-il Shin, Paul Wiegmann, Laurence Yaffe, and Martin Zwierlein for useful discussions.   AL is supported by the Smith Family Graduate Science and Engineering Fellowship, and would like to thank the Perimeter Institute for Theoretical Physics for hospitality when this work was initiated.  PC is supported by the Fundamental Laws Initiative of the Center for the Fundamental Laws of Nature at Harvard University.


\bibliography{refs}

\end{document}